\documentclass[aps,reprint,floatfix,superscriptaddress]{revtex4-2}
\usepackage[english]{babel}
\usepackage{epsfig}
\usepackage{subcaption}
\usepackage{amsmath}
\usepackage{enumerate}
\usepackage{bbold}
\usepackage{comment}
\usepackage{hyperref}

\usepackage{tikz}

\usepackage{amsmath,amsfonts,bm,mathrsfs,dsfont,amssymb,xfrac,bigints,cancel,mathtools,braket}

\usepackage{soul}

\newcommand{\pro}[1]{ {\rm pr}^{(#1)}}
\renewcommand{\Vec}[1]{\mathbf{#1}}
\renewcommand{\vec}[1]{{\bm{#1}}}

\newcommand{\dd}{\mathrm{d}}
\newcommand{\kk}{\vec{k}}
\newcommand{\uu}{\vec{u}}
\newcommand{\pp}{\vec{p}}

\newcommand{\grad}{\vec{\nabla}}
\newcommand{\diver}{\vec{\nabla}\!\cdot}

\newcommand{\pderiv}[2]{\dfrac{\partial#1}{\partial#2}}
\newcommand{\ptderiv}[1]{\pderiv{#1}{t}}

\DeclareMathOperator{\Li}{Li}

\begin{document}
	\title{Hydrodynamics in generalized electronic two-band systems}
	\author{E. Di Salvo}
    \affiliation{Institute for Theoretical Physics, Utrecht University, 3584CC Utrecht, The Netherlands.}
    \author{P. Cosme}
    \affiliation{Van der Waals--Zeeman Institute, Institute of Physics, University of Amsterdam,  Science Park 904, 1098XH Amsterdam, the Netherlands}
    \author{L. Fritz}
    \affiliation{Institute for Theoretical Physics, Utrecht University, 3584CC Utrecht, The Netherlands.}
	\begin{abstract}
		In this paper, we derive the Euler and Navier-Stokes equations for electronic two-band systems in arbitrary dimension and with generic power-law dispersion relations.
        We focus on the hydrodynamic transport regime, where such systems offer a unique tunability between a Fermi-liquid type regime at high doping and the inherent two-band physics of the low-density system close to the Dirac-type band-touching point. For a generic dispersion, the absence of Euclidean or Lorentzian invariance leads to novel types of hydrodynamic equations.
        We characterize these novel hydrodynamic regimes through dimensionless numbers, such as the Prandtl and Lorenz numbers, or the ratio between shear viscosity and entropy density.
		In all cases, we provide a derivation of the physics of the long-wavelength plasmonic modes. 
	\end{abstract}
	\maketitle

\section{Introduction}
\label{Introduction}

In recent years, the concept of hydrodynamics has been extended to electronic systems, where the collective flow of electrons in certain materials can mimic the behavior of (classical) fluids. This phenomenon, known as electronic hydrodynamics, occurs in systems where electron-electron interaction dominates over scattering from impurities, phonons, or other sources of relaxation \cite{lucas2018hydrodynamics, Fritz2023, Narozhny2022}. Under these conditions, the electrons behave as a strongly interacting fluid, characterized by collective motion rather than diffusive behavior of individual electrons.

Electronic hydrodynamics is distinct from the standard descriptions of diffusive charge transport. There the motion of individual electrons is hindered by scattering events with impurities or lattice vibrations~\cite{Mahan-90,FetterWalecka-71,Ziman2000}. Hydrodynamics is the description of systems close to
the equilibrium that are constrained in their evolution by conservation laws which are violated by impurities and lattice vibrations. An approximate hydrodynamic regime is then one where the mean free path for electron-electron scattering is much shorter than the mean free path for electron-impurity or electron-phonon scattering, allowing the electron fluid to exhibit collective behaviors such as vortices, Poiseuille flow, and in general viscosity-dominated dynamics \cite{gurzhi1968hydrodynamic}.

A classical fluid obeys the Navier-Stokes equation given by
\begin{equation}
n m \left( \partial_t \uu + \uu \cdot \nabla \uu \right) = -\nabla P + \eta \nabla^2 \uu + \mathbf{F}~,
\end{equation}
where $\uu$ describes the flow profile, $n$ is the density, $m$ is the particle mass, $P$ the pressure, $\mathbf{F}$ a force, and $\eta$ is the viscosity. For parabolic dispersion, the electronic analog follows the same equation~\cite{Zaanen2016} with the corresponding replacements. For a generic electronic system, one has to account for the quantum nature of the underlying densities, pressure, mass taking into account specifics of the dispersion relation. Consequently, the 'Navier-Stokes' equation has to be derived on a case-by-case basis.

The last years have witnessed an extraordinary production of novel materials with intrinsic quantum behavior, dubbed quantum materials~\cite{KeimerMoore}.
The advent of graphene, an atom-thin layer of carbon atoms arranged in a honeycomb lattice, lead to a plethora of novel systems with interesting electronic features, such as unusual electronic dispersions and generalized Dirac points~\cite{KeimerMoore,neto2009electronic}.
Graphene (and some of its versions discussed below) itself has emerged as one of the frontrunners for electronic hydrodynamics\cite{lucas2018hydrodynamics}. Due to its unique electronic structure, charge carriers behave as massless Dirac fermions~\cite{neto2009electronic}, graphene exhibits ultrahigh mobility, making it an ideal candidate for exploring non-conventional transport regimes \cite{bandurin2016negative}. At temperatures where electron-electron scattering dominates, the electrons in graphene flow as a viscous fluid, demonstrating behaviors such as negative local resistance, hydrodynamic whirlpools, and a violation of the Wiedemann-Franz law, which typically relates electrical and thermal conductivity in metals \cite{crossno2016observation}.
The stacking of different layers of graphene, for example, leads to novel types of electronic behavior~\cite{McCann2013TheGraphene}.
When two graphene sheets are stacked in a specific way, the new unit cell contains not just two but four atoms.
Close to the Fermi surface, the main excitations can be described as massive chiral quasiparticles with dispersion~\cite{McCann2013TheGraphene}
	\begin{equation}
		\label{BLDispRel}
		\omega_{\pm} =  \pm\frac{k^2}{2v}\;,
	\end{equation}
in sharp contrast to the case of monolayer, where chiral massless quasiparticles with linear dispersion are expected.
Chirality is, on the other hand, preserved by the stacking procedure (named Bernal stacking) that does not break inversion symmetry of the underlying lattice.
Iterating this procedure to three layers, one can stack \textit{à la} Bernal and get a dispersion relation that contains a linear and two parabolic subdispersions, or one can stack with the rhombohedral order~\cite{Zhang2010} and we get three different subbands that all disperse like
	\begin{equation}
		\label{TLDispRel}
		\omega_{i} =  \gamma_i k^3
	\end{equation}
and many other solutions are possible for graphene trilayers.
Graphene multilayered systems do not exhaust the list of quantum materials: many other systems, like Weyl semimetals, media with excitonic or polaronic excitations or even strange metals, populate it~\cite{KeimerMoore}.

Transport properties, such as electric or thermal conductivity, are intrinsically contained in the hydrodynamic description but their validity is intimately related to the validity of the hydrodynamic approximation itself. Additionally, there are viscosity related phenomena such as the formation of vorticity~\cite{Levitov2016}. The viscosity in an electron fluid can be probed experimentally using techniques such as nonlocal transport measurements and imaging of current flow patterns, which reveal deviations from diffusive behavior \cite{bandurin2016negative}. These insights can in principle open the door to the engineering of novel devices where hydrodynamic effects are harnessed for applications in low-dissipation electronics, quantum information technologies, and efficient energy transport.

This work aims to derive the hydrodynamic equations of generic electronic systems and characterize their transport properties in this regime.
We derive the hydrodynamic equations from the kinetic theory, both in the ideal and in the viscous regime.
Following up on that, we derive the electric and thermal conductivities and we study the Wiedemann-Franz ratio in those systems.
Finally we also derive the plasmon dispersion relation induced by Coulomb interaction.
	
We consider electrons with dispersion relation
\begin{equation}
\label{eq:powerlaw}
    \omega_\pm = \pm B k^a \mbox{,  } a\geq0\mbox{ , }B>0
\end{equation}
which is characteristic for a low-energy description of fermionic excitations in the vicinity of band touching points.
In equilibrium, the excitations obey the Fermi-Dirac distribution~\cite{Mahan-90}
\begin{equation}
    \label{DistFunc}
    f_\mathrm{eq}^{(\pm)}(\omega) = \pm\frac{z^{\pm1}}{e^{\omega/(k_BT)} + z^{\pm1}}~,
\end{equation}
where $T$ represents temperature and $z$ the fugacity.
To include the effects of spatial and temporal inhomogeneities we assume $f(\vec{x},\kk,t)$~\cite{landau1987fluid}. In the presence of external forces $\vec{F}$, this function obeys the Boltzmann equation
\begin{equation}
    \label{BoltzEq}
    \partial_t f^{(\pm)} + \vec{\nabla}_{\kk}\omega_\pm\cdot\vec{\nabla}_{\vec{x}}f^{(\pm)} + \vec{F}_\pm\cdot\vec{\nabla}_{\kk}f^{(\pm)} = \mathcal{C_\pm}[f^{(+)},f^{(-)}] ,
\end{equation}
where the presence of the scattering integral $\mathcal{C_\pm}[f^{(+)},f^{(-)}]$ accounts for Coulomb interaction between charged particles and is in general a very complicated object.
In the present form, we neglect other scattering effects such as disorder and/or phonon coupling. The underlying assumption is that Coulomb interaction is the dominant relaxational mechanism. 
The above equation provides the starting point for the derivation of the hydrodynamic equations as well as for the study of transport properties. 

The paper is structured as follows: In Sec.~\ref{Thermodynamics} we describe the equilibrium properties of the models we are investigating and we exploit them in Sec.~\ref{Conservation laws} to express conservation laws in Euler and Navier-Stokes regime without relying on the details contained in dispersion relation.
Such details matter only when prefactors are specified, but they do not change the generic form of the equations.
In Sec.~\ref{Transport coefficients and Lorentz number}, we characterize further the thermoelectric response of electronic systems beyond the hydrodynamic regime and in Sec.~\ref{Collective Modes} the emergence of collective hydrodynamic modes is presented.
We gather our conclusions and outlook in Sec.~\ref{Conclusions}.
    
\section{Thermodynamics}
\label{Thermodynamics}

We are in general concerned with a fluid close to equilibrium that has a small drift velocity $\vec{u}$. The corresponding distributions read 
\begin{equation}
\label{eq:DistFuncStream}
     f^{(\pm)}_\text{stream}(\vec{x},\kk,t) = \pm\frac{z^{\pm1}}{e^{(\omega - \kk\cdot\uu)/k_BT} + z^{\pm1}}~\;.
 \end{equation}
A key assumption is that the fluid velocity $\uu$ is small compared to the characteristic velocity of the electrons $v_F = aB[\mu + (d/a)k_BT]^{a-1}$. This allow us to expand the functions to lowest order in the stream velocity according to 
 \begin{equation}
     f^{(\pm)}_\text{stream}=f^{(\pm)}_\mathrm{eq}+f^{(\pm)}_\mathrm{eq}\left[1-f_\mathrm{eq}^{(\pm)}\right]\frac{\vec{u}\cdot\kk}{k_BT} +\mathcal{O}\left(\frac{u^2}{v_F^2}\right)~\;.
 \end{equation}

In local equilibrium we can use the equilibrium distribution functions for electrons and holes \eqref{DistFunc} to calculate equilibrium thermodynamic quantities~\cite{landau1987fluid,Fritz2023}, meaning we can neglect the drift. In the following we work in $d$ dimensions ($d>1$). %
For the thermodynamic densities, we find %
\begin{multline}
    n
    = 
    -\frac{2^{1-d}\pi^{-\frac{d}{2}}}{aB^{\frac{d}{a}}\Gamma\left(\frac{d}{2}\right)}(k_BT)^{\frac{d}{a}}\Gamma\left(\frac{d}{a}\right)\times\\
    \left[\Li_{\frac{d}{a}}\left(-z\right) - \Li_{\frac{d}{a}}\left(-z^{-1}\right) \right],
\end{multline}
for the charge density, and 
\begin{multline}
        n\langle E\rangle
        =-\frac{2^{1-d}\pi^{-\frac{d}{2}}}{aB^{\frac{d}{a}+1}\Gamma\left(\frac{d}{2}\right)}(k_BT)^{\frac{d}{a}+1}\Gamma\left(\frac{d}{a}+1\right)\times\\
        \left[ \Li_{\frac{d}{a}+1}\left(-z\right) + \Li_{\frac{d}{a}+1}\left(-z^{-1}\right) \right]
\end{multline}
for the internal energy density (note that we took $\hbar=1$) and we are expressing the integrals as polylogarithm functions $\Li_s$~\cite{1964HandbookTables}.

The average momentum  $\langle\vec{p}\rangle$ is zero in the absence of drift, so we have to go to linear order in the drift.
In this regime the Galilean symmetry is effectively restored for the distribution function of the system, since its form \eqref{eq:DistFuncStream} does not change after a boost of velocity $\delta\vec{v}$, given that corrections to $\omega - \kk\cdot\uu$ are of orders higher than linear. 
From the Boltzmann equation perspective, this can be explained as the collisional invariants are also changed accordingly by the action of the boost: then the distribution function before and after the boost still annihilates the collision operator of \eqref{BoltzEq}, hence it represents a local equilibrium state also after such a boost.
For small boosts, the change in a particle's momentum can be linearized and it reads $\delta\kk = k^{2-a}/[a(a-1)B]\delta\vec{v}$; in a fully Galilean invariant system, the shift of particle's momentum does not depend on the momentum in the system at rest, hence only in the $a=2$ case Galilean invariance of the equilibrium distribution is recovered also at higher orders.
In addition to it, the left-hand side of \eqref{BoltzEq} is modified by a term $[(a-2)/(a-1)]\delta\vec{v}\cdot\nabla_\vec{x}f^{(\pm)} - \{(a-2)/[a(a-1)B]\}k^{a-4}(\vec{F_\pm}\cdot\kk)\delta\vec{v}\cdot\nabla_\kk f^{(\pm)}$, which is a first order correction in $\delta\vec{v}$ that vanishes exactly for $a=2$; in this case also the space and time evolution of the distribution is invariant for Galilean observers.

Eventually, the momentum density reads 
\begin{multline}
        n\langle\vec{p}\rangle%
    =-\vec{u} \,\frac{2^{1-d} \pi ^{-d/2}(d+2-a)}{a^2 B^{(d+2)/a}\Gamma\left(\frac{d}{2}\right) }(k_BT)^{\frac{d+2}{a}-1}\times\\
    \Gamma\left(\frac{d+2}{a}-1\right)\left[\Li_{\frac{d+2}{a}-1}\left(-z\right) - \Li_{\frac{d+2}{a}-1}\left(-z^{-1}\right) \right]. 
\end{multline}

We can define the hydrodynamic mass element as $m^\star n\uu = n\langle\pp\rangle$, {\it i.e.},
\begin{multline}
    m^\star=\frac{(d+2-a)}{2aB^{2/a}}\frac{ \Gamma\left(\frac{d+2}{a}-1\right)}{  
    \Gamma\left(\frac{d}{a}\right)}(k_BT)^{\frac{2}{a}-1}\times\\
    \frac{\Li_{\frac{d+2}{a}-1}\left(-z\right)-\Li_{\frac{d+2}{a}-1}\left(-z^{-1}\right)}{\Li_{\frac{d}{a}}\left(-z\right) - \Li_{\frac{d}{a}}\left(-z^{-1}\right)}~.
    \label{eq:hydromassgeneral}
\end{multline}
Defining this quantity is crucial for the treatment of a generic hydrodynamic theory. It connects the continuity equation (which involves $n$ and $\vec{j}=n\vec{u}$) to the equation of momentum transport (which contains both $\vec{u}$ and $n\langle\vec{p}\rangle$). While for parabolic systems the microscopic effective mass $m_\text{eff}=1/(\partial_{k}^2E)$ is well-defined and constant, meaning that $\langle m_\text{eff}\rangle=m_\text{eff}=m^\star$, for a generic energy relation like Eq.\eqref{eq:powerlaw} this is not the case and in general $\langle m_\text{eff}\rangle\neq m^\star$ as expressed by Eq. \eqref{eq:hydromassgeneral} as $m^\star = 1/(2B)=\text{const}=m_ \text{eff}$ if, and only if, $\alpha=2$, irrespective of the dimension $d$. The special case of Dirac materials with the linear $E=v_F |\vec{k}|$ presents an even more drastic situation as the effective mass diverges, yet this definition of the mass element circumvents the issue and even evinces that, for such ultra-relativistic dispersion, the mass is determined by the energy alone as
\begin{equation}
    m^\star=\frac{d+1}{2}\frac{\langle E\rangle}{B^2}=\frac{d+1}{2}\frac{\langle E\rangle }{v_F^2}~,\label{eq:hydromass_dirac}
\end{equation}
coupling the energy and momentum transport~\cite{Pongsangangan2023TheGraphene}. Regrettably, these two specific cases of $a=2$ and $a=1$ are the only ones that allow bringing together the transport equation of the macroscopic quantities that reflect the microscopic ones conserved at the collisions for any $\mu/(k_BT)$. That is the only way to match the order of the polylogarithmic part of the equilibrium functions.

\section{Conservation laws}
\label{Conservation laws}
\subsection{Transport equations}
\label{Transport equations}
The general procedure to obtain a hydrodynamic theory from the kinetic theory is to multiply the Boltzmann equation with successive moments in $\vec{p}^n$ and integrate over it. In that way $n=0$ corresponds to the conservation of total charge, $n=1$ to the momentum density, and the subsequent ones translate into transport equations for total energy density or related quantities~\cite{Chapman1954TheGases,landau1987fluid}.
The integral over the incoming momentum of the collision integral is identically null, {\it {i.e.}} $\int\mathcal{C}_\pm\,{\rm d\vec{p}}= 0$, guaranteeing the conservation of the total charge~\footnote{The situation changes when there are different species of particles involved that can decay into each other or are involved in emission/absorption processes, we address that situation later.}.
For higher order moments that relate to other collisional invariants, such as momentum and/or energy, it holds that $\int \vec{p}^j\left(\mathcal{C}_+ + \mathcal{C}_-\right)\,{\rm d}\vec{p}= 0$ for $j\neq0$ (we keep $j$ generic here to accommodate for different forms of dispersion relations). Integrating the kinetic equation (\ref{BoltzEq}) over the moments $\vec{p}^j$~\cite{Vlasov68} %
leads to the conservation equations for total charge, momentum and energy
\begin{subequations}
\label{Euler}
    \begin{equation}
            \partial_t n+\diver(n\vec{u}) = 0~,
    \end{equation}
    \begin{equation}
        \partial_t  n \langle \vec{p}\rangle+\diver \left[n  \langle \vec{p}\rangle\otimes\vec{u} +{\vec{\Pi}}\right]  -n\vec{F}=0~,
    \end{equation}
    \begin{equation}
        \partial_t n \langle E\rangle+\diver \left[
n\langle  E\rangle \vec{u} +{\vec{Q}} \right]  -n\vec{F}\cdot\vec{u}=0~.
    \end{equation}\label{eq:transport_general}%
\end{subequations}%
Up until now, no details about the specific system were required and Eqs. \eqref{eq:transport_general} are formal and universal. To progress further, we need information about the distribution function and energy dispersion in order to evaluate the pressure tensor and heat current and, more importantly, to relate average momentum and average velocity.
When the generic dispersion relation \eqref{eq:powerlaw} is considered, we can derive the exact form of scalar pressure
\begin{equation}
		\label{PressDef}
		P = \frac{aB}{d}\int_{\mathbb{R}^d}\frac{\dd^d \kk}{(2\pi)^d} k^a \left[f^{(+)}_\mathrm{eq}(\omega) - f^{(-)}_\mathrm{eq}(\omega)\right] = \frac{a}{d}n \langle E\rangle,
	\end{equation}
and heat current
\begin{equation}
    \label{HeatCurr}
    \vec{Q} = \frac{a}{d}n \langle E\rangle\vec{u} = P\vec{u}~,
\end{equation}
which are linked via the definition of the total energy of the system; usually this relation is introduced by the means of defining the total enthalpy of the system $W = n \langle E\rangle + P$ and the total energy current is given by $\vec{j}^E = W\vec{u}$~\cite{Narozhny2022}.
The scalar pressure is connected to the pressure tensor as $\vec{\Pi} = P\vec{1}$, where $\vec{1}$ stands for the identity tensor.

Indeed, it is clear that Eqs. \eqref{eq:transport_general} are not closed in general, even discarding the $\vec{\Pi}$ and $\vec{Q}$ currents. The only two aforementioned cases where there is a natural closure are exactly the parabolic bands of $a=2$ and the Dirac bands with $a=1$. In the former, average momentum and velocity are naturally paired and $n \langle \vec{p}\rangle\propto n  \vec{u}$~\cite{Vlasov68}; the constant of proportionality can be identified as the effective (or dressed) mass of the model $m^*$.For the latter, the linear dispersion (i.e., usual graphene dispersion in 2D) makes the effective mass vanishe but, as a consequence, one retrieves the expected $n\langle \vec{p}\rangle \propto n\langle E\rangle\uu$~\cite{Pongsangangan2023TheGraphene,Narozhny2021HydrodynamicGraphene}. However, for an arbitrary band, apart from these two cases, there is no natural way to close the problem. So, one needs to resort to the definition of the hydrodynamic mass of Eq. \eqref{eq:hydromassgeneral}.

In the Dirac limit $\mu/k_BT\to 0$, the total charge of the system vanishes per definition since we are at charge neutrality.
While the hydrodynamic mass (\ref{eq:hydromassgeneral}) remains finite and independent from the chemical potential, the momentum is forced to be zero, too. The only non vanishing quantity at charge neutrality is correctly the total energy.
Hence we are forced to study small charge and momentum fluctuations in the vicinity of the Dirac point, generated by fluctuations of the chemical potential; we are anyhow able to show that all of them are related to the total energy density properties at charge neutrality.
We can rewrite all the quantities as thermal fluctuations given by linear terms of $\mu/(k_BT)$
\begin{subequations}
    \begin{align}
        n&\xrightarrow{k_BT/\mu\to\infty} -\frac{2^{2-d}\pi^{-\frac{d}{2}}}{aB^{d/a}\Gamma\left(\frac{d}{2}\right)}\Gamma\left(\frac{d}{a}\right)\zeta\left(\frac{d}{a}\right)\nonumber\\
        &\times(k_BT)^{\frac{d}{a}}\frac{\mu}{k_BT}~,\\
        n\langle E\rangle&\xrightarrow{k_BT/\mu\to\infty}-\frac{2^{2-d}\pi^{-\frac{d}{2}}}{aB^{d/a}\Gamma\left(\frac{d}{2}\right)}\Gamma\left(\frac{d}{a}+1\right)\zeta\left(\frac{d}{a}+1\right)\nonumber\\
        &\times(k_BT)^{\frac{d}{a}+1}~ ,\\
        n\langle \pp &\rangle\xrightarrow{k_BT/\mu\to\infty} -\vec{u}\,\frac{2^{2-d} \pi ^{-d/2}(d+2-a)}{a^2 B^{(d+2)/a}\Gamma\left(\frac{d}{2}\right)}\\
        &\times\Gamma\left(\frac{d+2}{a}-1\right)\zeta\left(\frac{d+2}{a}-1\right)(k_BT)^{\frac{d+2}{a}-1}\frac{\mu}{k_BT},\nonumber\\
        m^*&\xrightarrow{k_BT/\mu\to\infty}\frac{(d+2-a)}{2aB^{2/a}}\frac{\Gamma\left(\frac{d+2}{a}-1\right)}{\Gamma\left(\frac{d}{a}\right)}\frac{\zeta\left(\frac{d+2}{a}-1\right)}{\zeta\left(\frac{d}{a}\right)}\nonumber\\
        &\times(k_BT)^{\frac{2}{a}-1},
    \end{align}
\end{subequations}
where $\zeta(s)$ is the Riemann zeta function.
Since the hydrodynamic mass is well defined in this limit, we can close the relation between energy and momentum (we disregard fluctuations in the total charge density since they are $n\simeq \left[\mu/(k_BT)\right]\delta n$).
The system of equations reduces to
\begin{subequations}
    \begin{align}
        \partial_t\uu + (\uu\cdot\nabla)\uu + \frac{1-a}{d}(\diver\uu)\uu + \frac{1}{m^*}\grad [n\langle E\rangle] = 0 ~, \\
        \partial_t n \langle E\rangle + \left(1+\frac{a}{d}\right)\diver \left[n\langle  E\rangle \vec{u}\right]=0~,
    \end{align}
\end{subequations}
where the force terms are negligible in the $\mu/(k_BT)$ expansion since they are multiplied by the particle density $n$.

All the previous quantities also simplify in the Fermi liquid limit, $\mu/k_B T\to \infty$. There, $-\Li_s(\pm e^{\mu/k_BT})\to \frac{(\mu/k_BT)^s}{\Gamma(s+1)}$ and so the densities become
\begin{subequations}
\begin{align}
    n&\xrightarrow{\mu/k_BT\to\infty}\frac{2^{1-d}\pi^{-\frac{d}{2}}}{dB^{d/a}\Gamma\left(\frac{d}{2}\right)} \mu^{\frac{d}{a}},
\\
        n\langle E\rangle&\xrightarrow{\mu/k_BT\to\infty}\frac{2^{1-d}\pi^{-\frac{d}{2}}}{(d+a)B^{d/a}\Gamma\left(\frac{d}{2}\right)} \mu^{1+\frac{d}{a}},
\\
    n\langle\vec{p}\rangle &\xrightarrow{\mu/k_BT\to\infty}\vec{u}\frac{2^{-d} \pi ^{-d/2}}{a B^{(d+2)/a}\Gamma\left(\frac{d}{2}\right) }\mu^{\frac{d+2}{a}-1} , 
\\    m^\star&\xrightarrow{\mu/k_BT\to\infty}
    \frac{d\mu^{\frac{2}{a}-1}}{ 2aB^{2/a}}~.
\end{align}
\end{subequations}
We can now reexpress the hydrodynamic mass with the density according to
\begin{equation}
    m^\star \propto n^{\frac{2-a}{d}},
\end{equation}
and equivalently the equation of state
\begin{equation}
    P \propto n^{\frac{d+a}{d}}.
\end{equation}
Within this approach where the thermal effects are no longer relevant, and the pressure is given by an equation of state, the energy equation loses significance and the continuity and momentum equation are enough to describe the dynamics of the fluid. The continuity equation is not altered in any way by the particularities of dimensionality or band structure because it governs an elemental property of the charges; the momentum equation however is modified when written in terms of the velocity field. From the previous $m^\star=m^\star_0 n^\frac{2-\alpha}{d}$ and taking $\Pi_{ij}=P\delta_{ij}=P_0n^\frac{d+\alpha}{d}\delta_{ij}$ while using also the continuity equation to eliminate $\partial_t n$ terms we arrive at 
\begin{eqnarray}
      \partial_t \vec{u}+(\vec{u}\cdot\grad)\vec{u}-\frac{2-a}{d} \vec{u}(\diver\vec{u})\nonumber\\
      + \frac{(d+a) P_0}{dm^\star_0 }n^{\frac{2(a-1)}{d}-1}\grad n -\frac{\vec{F}}{m^\star}=0,\label{eq:generalmomentum}
\end{eqnarray}
thus, for $a\neq 2$ the convective part has an additional term which, as we will see shortly, changes the symmetry properties. 

One consequence of the variable mass element is the loss of boost invariance, i.e. the system is neither Galilean nor Lorentz invariant. Let us see why this is the case. We established in Eq.\eqref{eq:generalmomentum} that, for a variable hydrodynamic mass element (following a power law), the convective part of the momentum equation acquires an additional term $\sim \vec{u}\diver\vec{u}$.
It can be shown (see Appendix \ref{Appendix: Breaking of Galilean invariance}) that such a non-linear contribution breaks any boost-like symmetry; that is, the Lie algebra of \eqref{eq:generalmomentum} only includes the generator of boosts for $a=2$.
This is however the consequence of considering a microscopic description that already violates Galilean invariance, due to the dispersion relation of the single excitations.
This eventually reflects in the coarse grained theory upon the definition of the hydrodynamic mass element and becomes more evident when we shift our focus to the conserved current associated with the boosts.
It is known that the Galilean symmetry implies the conservation of the center-of-mass movement, since in the center-of-mass frame the whole kinetic energy is given by the motion of a particle with center-of-mass velocity and mass equal to the sum of all the masses in the system.
But suppose that a collection of particles undergoes a uniform boost in momentum space, even though the sum of their momenta transforms accordingly, the center-of-mass velocity acquires extra contributions which are just a term proportional to the center-of-mass velocity and the kinetic energy of the center-of-mass moving with velocity of the boost.
However, for (\ref{BLDispRel}) with $a\neq 2$, this does not hold at all and it is also impossible defining the kinetic energy in center-of-mass frame only in terms of center-of-mass coordinates.

\begin{figure}[ht!]
    \centering
    \includegraphics[width=0.85\linewidth]{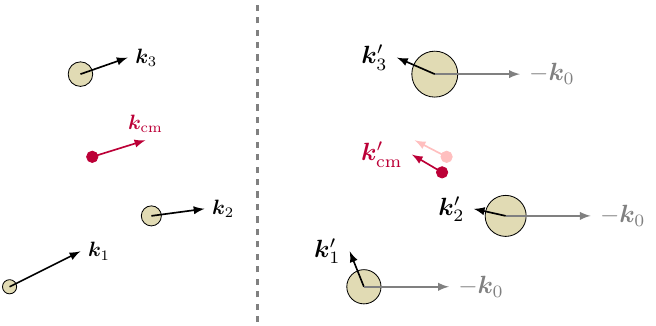}
    \caption{Schematic illustration of broken boost invariance. Considering a collection of particles which undergo a uniform boost $k_0$ in momentum space. With a non quadratic band structure, the energies/masses (here represented by the radii of the disks) break the boost invariance of the movement of the centre of mass $k'_\text{cm}$-- dark red vs. light red on the right pannel.}
    \label{fig:nonGalileanboost}
\end{figure}

As such, it is now evident that even in the case of relativistic particles, i.e. with a linear dispersion as graphene electrons, the coarse grained hydrodynamic description does not inherit the Lorentz group structure. That is, the invariant action of boosts is not possible for $a\neq2$ since $ m^\star\neq \rm{const.}$ 
Moreover, any transformation of the mass element, potentially emergent from the band structure of the charge carriers, has a limited influence on the overall behaviour of the system. 
Given our definition of hydrodynamic mass, a boost would, at most, amount to second-order corrections $m^\star=m^\star_0+\mathcal{O}(\delta v^2/v_F^2)$. In real devices, the drift velocity of electrons saturates to values well below the Fermi velocity; thus, the modelling of nanoelectronic devices such as graphene transistors and similar structures can safely disregard $\delta v^2/v_F^2$ corrections. Moreover, in the case of Dirac fermions, one can always circumvent the hydrodynamic mass problem using the exact relation $\langle\vec{p}\rangle=3\vec{u}\langle E\rangle/2v_F^2$ at the cost of adding complexity.

Having derived the Euler equations for our system, it is possible to find the sound waves in the fluid by describing transport equation for the temperature at lowest order in $\uu$, where Galilean invariance is effectively retrieved.
We relate the temperature to the internal energy of the system by relying on the equipartition of energy $\langle U\rangle = (d/a)g(\mu,T)k_BT$.
It follows from thermodynamic relations considering that internal energy is independent of the choice of reference frame and we can write it as a linear combination $n\langle U\rangle = n\langle E\rangle - n\langle \pp\rangle\cdot\uu$~\cite{landau1987fluid}.
The function $g(\mu,T)$ contains only negative power-law or exponentially decaying corrections to equipartition.
By doing so we get the following equation for temperature at lowest order
\begin{equation}
    \partial_t T + \uu\cdot\grad T + \frac{a}{d}\diver\uu=0~.
\end{equation}
By linearizing the system and employing Fourier transformation, speed of sound in the liquid reads
\begin{equation}
    \label{SoundVelocity}
    v_s = \sqrt{\frac{k_BT}{2m^*}\left(1+\frac{a}{d}\right)}~.
\end{equation}
This result allow us to identify the specific heat at fixed volume $c_V = (d/a)k_B$ and pressure $c_P = (1+d/a)k_B$ of the system, which is exactly the result predicted by equipartition of energy for ideal fluids.

\subsection{Beyond conservation}
\label{Beyond conservation}
The integration of the collisionless Boltzmann equation leads us to transport equations, akin to the Euler equations, that explicitly conserve charge, momentum, and energy, and we argued that the collisionless limit was adequate provided that the system is close enough to equilibrium. However, we know that the hyperbolic system of transport equations is just the first-order approximation in gradient expansion\footnote{Even in classical fluid theory one goes beyond Euler to Navier--Stokes introducing parabolic operators, i.e. diffusive.}. In order to go further, one needs to consider the diffusion of the quasi-conserved quantities.   

Indeed, the collisional operators have not yet been discussed, aside from stating that they drive the corresponding distributions towards equilibrium. By construction, those terms are irrelevant for the continuity equations as they inherently conserve the number of particles but they can diffuse momentum and energy\footnote{As it becomes clear when we progress from Boltzmann to Fokker--Planck.}. To go further, one needs to uncover the differential operators that make the fluid theory, potentially, nonconserving. When deriving the fluid equations, in a completely general approach we had to define the viscous tensor and heat current as $ \vec{\widetilde \Pi}^{(\pm)}=\int_{\vec{p}}(\vec{p}-\langle\vec{p}\rangle)\otimes(\vec{v}_\pm-\vec{u})f^{(\pm)}$ and $ \vec{\widetilde Q}^{(\pm)}=\int_{\vec{p}}(E-\langle E \rangle)\otimes(\vec{v}_\pm-\vec{u})f^{(\pm)}$, respectively, and at that point we restricted ourselves to calculate the static contribution, i.e., integrating over $f^{(\pm)}_\mathrm{stream}$. The integration over the perturbations to the distribution function provides the intended parabolic corrections.    
To achieve this goal, we use the Chapman--Enskog method \cite{Esposito1999,chapman1990mathematical} (up to second-order) and obtain the diffusive description of the fluid.
In such a context, we allow the collision operator to relax towards the equilibrium with a certain relaxation timescale
\begin{equation}
    \mathcal{C}_\pm[f^+,f^-] = \mp\left[\frac{f^{(+)}}{\tau_+} - \frac{f^{(-)}}{\tau_-}\right]~,
\end{equation}
disregarding the effects of disorder or external phononic baths.
At this point, it is important to note that such a timescale impacts the transport of momentum (leading to viscosity) and energy (which translates to heat diffusivity), whereas the continuity equation remains unaffected: The total number of particles is assumed to be conserved.
Nonetheless, on general grounds, the density transport can also gain a diffusive term when considering the lifetime of the quasiparticles, but note that such timescale is not linked to the relaxation time of the binary collisions in the systems we are considering.
Also, for a hydrodynamic theory to hold, the particles must be sufficiently long-lived which renders the breaking of particle number conservation irrelevant~\cite{Fritz2023}. 

At this point, we can specify bulk and shear viscosity induced by the generic dispersion relation \eqref{eq:powerlaw}: To do it, we write the streaming term as a function of the inhomogeneous and slowly varying chemical potential, temperature and fluid velocity~\cite{landau1987fluid}.
We eliminate the time derivatives the Euler equations \eqref{Euler} and by introducing the relative velocity $\vec{c}_\pm = \vec{v}_\pm-\uu$ while expressing the dispersion relation via particle velocity as $\omega_\pm = (1/a)\kk\cdot\vec{v}_\pm$ we get the following form:
\begin{multline}
\mathfrak{D}f^{(\pm)}_\mathrm{stream}=
    f^{(\pm)}_\mathrm{stream}\left[1-f^{(\pm)}_\mathrm{stream}\right]\\ \times\Bigg\{%
    \vec{c}_\pm\cdot \grad T\left(\frac{\left(\frac{1}{a}\vec{v}_\pm - \uu\right)\cdot\vec{k}}{T^2}+s\right)-\vec{c}_\pm\cdot \grad P\\
    +\frac{\vec{k}}{T}\cdot\left\{(\vec{c}_\pm \cdot\grad)\vec{u}-\left[\frac{1}{d}\vec{v}_\pm - \left(\frac{a}{d}+1\right)\uu\right]\grad\cdot \vec{u}\right\}
    \Bigg\}~.
\end{multline}
That enters the expression of the correction to the stress tensor according to
\begin{equation}
    \vec{\widetilde{\Pi}} = -\int\frac{\dd^d\mathbf{k}}{(2\pi)^d}\mathbf{k}\otimes\left[\tau_+\vec{v}_+\mathfrak{D}f^{(+)}_\mathrm{stream} - \tau_-\vec{v}_-\mathfrak{D}f^{(-)}_\mathrm{stream}\right].
\end{equation}
After integration and balancing pressure terms, the previous expression implies vanishing bulk viscosity
\begin{equation}
    \label{BulkViscosity}
    \eta_B = 0~, 
\end{equation}
which is expected using the dilute gas approximation.
However, for the shear viscosity we need to distinguish three different cases in order to carry out computations: the bi-dimensional, tri-dimensional, and higher dimensional one.
The results for the last one are shown in Appendix \ref{Appendix: Derivation of shear viscosity}.
The reason behind this distinction, and why it does not happen for other quantities investigated beforehand, is that shear viscosity pairs different components of momentum and it is not related to an isotropic rank-two tensor, but to a rank-four tensor (see also Appendix~\ref{Appendix: Integration over isotropic multi-dimensional momentum space} for more details).
For $d=2$ we get
\begin{equation}
    \label{ShearViscosity2D}
    \eta_S^{2D} = \left(\frac{a+2}{4}\right) (\tau_+P_++\tau_-P_-)~,
\end{equation}
and for $d=3$ we would find
\begin{equation}
    \label{ShearViscosity3D}
    \eta_S^{3D} = \left(\frac{a+3}{5}\right) (\tau_+P_++\tau_-P_-)~,
\end{equation}
where the partial pressure of the species is labeled by $P_\pm$.
This quantity is well defined in the clean limit.
Both results agree with those for graphene ($d=2$ and $a=1$)~\cite{Pongsangangan2023TheGraphene} and a dilute gas ($d=3$ and $a=2$)~\cite{Vlasov68}. Microscopic details are encoded in the relaxation time.

Another way to express this quantity is considering the ratio with entropy density $s(T,\mu) = (W-n\mu)/T$ at the equilibrium
\begin{equation}
    \label{ViscosityVsEntropy}
    \frac{\eta^{d}_S}{s} = \mathcal{G}_d(a)(\tau_+P_++\tau_-P_-)\frac{T}{W-n\mu}~,
\end{equation}
where we stored the dimensional dependence of the shear viscosity in a geometric prefactor $\mathcal{G}_d(a)$ and expressed entropy density in terms of entalphy (which is in turn proportional to energy) and charge density.
Both energy and entropy times the inverse relaxation times scale with the same power law in the vicinity of Dirac point, given that charge density just vanishes, and the ratio does not explicitly depend on the temperature and chemical potential.
This is in agreement with results for weakly interacting systems, where temperature affects only enters via renormalization of parameters of the theory as $B$~\cite{Fritz2008QuantumGraphene}.
It is not the case in the Fermi liquid limit, where the $\mu n$ contribution scales differently from energy and leads to a divergence of the ratio, especially for fairly small values of $T$, or conversely very large $\mu$.
It is then expected that such systems behave similar to perfect fluids at low temperatures in the vicinity of the Dirac point, while they deviate strongly away from it.

\begin{figure}
  \centering
  \begin{subfigure}{0.96\columnwidth}
    \includegraphics[width=\linewidth]{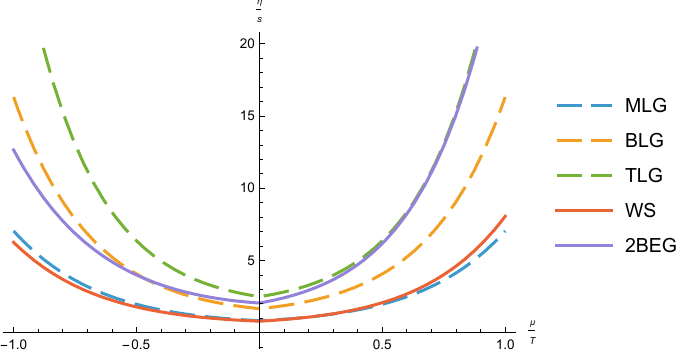}
    \caption{}
  \end{subfigure}
  \hfill
  \begin{subfigure}{0.96\columnwidth}
    \includegraphics[width=\linewidth]{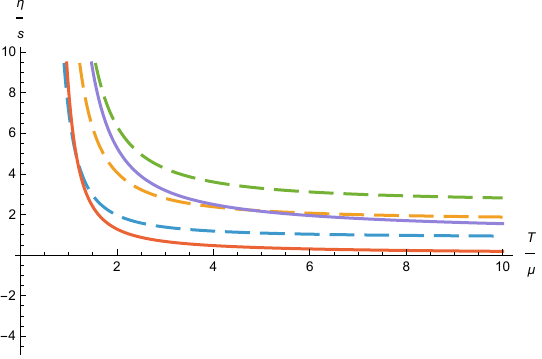}
    \caption{}
  \end{subfigure}
  \caption{
  We plotted here the value of the ratio between viscosity and entropy (\ref{ViscosityVsEntropy}) against the chemical potential in units of temperature $\mu/T$ (panel a) or temperature in units of chemical potential $T/\mu$ (panel b) for monolayer (MLG), bilayer (BLG) and trilayer graphene (TLG) in dashed lines, while in thick ones are represented the behaviours for Weyl semimetals (WS) or two-band electronic gases (2BEG). From both figures we can appreciate how, in the vicinity of Dirac point, effects of viscosity are minimal on transport properties; on the other hand, the Fermi liquid limit, even in absence of disorder, is characterized by strong dissipative contributions in the momentum sector. We employed a relaxation time as $\tau_0 = 0.6T\exp{(-|\mu|/T)}$, assuming Onsager's relation (\ref{RelTimeRelation}) to hold.
  }
  \label{Fig:Eta}
\end{figure}

It is worth to note that in the derivation of the viscosity, one finds a completely symmetric rank-four tensor --- hence fully characterized by one scalar, the shear viscosity --- because we considered only isotropic interactions and discarded any momentum dependence of the forces. If one considers a more general form or the force term on the kinetic Boltzmann equation, it can contribute to the viscous terms. This occurs, for instance, when dealing with magnetized systems where the viscous tensor displays an antisymmetric part, known as odd viscosity~\cite{Avron1998OddViscosity,fruchart_odd_2023}. However, in those cases, both the shear and odd viscosities scale with $\mathcal{G}_d(a)(\tau_++\tau_-)P$, sharing the same geometrical prefactor, making our results on the dimensionality and band exponent straightforward to apply.

\section{Transport coefficients and Lorentz number}
\label{Transport coefficients and Lorentz number}
Certain quantities, such as the shear viscosity, are well-defined ain the absence of disorder~\cite{Fritz2008QuantumGraphene}.
In one band systems, momentum relaxation is always ensured by the presence of a dissipating collisional operator~\cite{Vlasov68}.
This is not the case for two-band systems: At the Dirac point, particle-hole symmetry guarantuees that momentum dissipated by one band is directly transferred to particles in the other, and vice versa, preserving its total value~\cite{Pongsangangan2022ThermoelectricPairs}.
Hence, when we investigate transport quantities in such set-up we need to ensure momentum dissipation also in that case; the simplest way to do it is considering disorder, which couples with the two bands in the same fashion and avoids conservation of momentum also at the Dirac point~\cite{Pongsangangan2022HydrodynamicsTransport}.

We can now consider probing the two-band electronic system by applying an electric field $\vec{E}$ and a thermal gradient $\vec{\nabla}T$ in the two regimes (Dirac and Fermi liquids) we illustrated beforehand; in describing the linear response to this we mostly follow~\cite{Pongsangangan2022HydrodynamicsTransport}.
Provided that these perturbations are not strong enough to drive the system out of the equilibrium beyond linear response, the first order relaxation time approximation version of the Boltzmann equation still holds
\begin{equation}
	\label{BETransp}
    \begin{split}
	&- \frac{(\omega_\pm - \mu)}{T}\vec{\nabla}T\cdot\vec{\nabla}_{\kk} f^{(\pm)} -  e\vec{E}\cdot\vec{\nabla}_{\kk} f^{(\pm)} =\\ 
    & \qquad -\frac{\delta f^{(\pm)}}{\tau_{\pm}}+\frac{\delta f^{(\mp)}}{\tau_{\mp}}-\frac{\delta f^{(\pm)}}{\tau_\mathrm{dis}}~,
    \end{split}
\end{equation}
with modifications introduced by the electric field and the thermal gradient to the particle and hole distributions.
We only consider DC responses and therefore drop the time derivatives.
It is now straightforward to derive the charge and thermal currents $\delta \vec{j}$ and $\delta \vec{j}^Q$ by integration by solving the related linearized system~\cite{Pongsangangan2022HydrodynamicsTransport}. It leads to the Onsager matrix with the associated reciprocal relations
\begin{equation}
    \begin{pmatrix}
        \delta \vec{j}\\
        \delta \vec{j}_Q
    \end{pmatrix} = \mathbb{L}\begin{pmatrix}
        \vec{E}\\
        \frac{\vec{\nabla}T}{T}
    \end{pmatrix}~,
\end{equation}
in which the Onsager matrix reads
\begin{equation}
    \mathbb{L} = \begin{pmatrix}
        \sigma && \alpha \\
        \bar\alpha && \kappa
    \end{pmatrix}~.
\end{equation}
Then we can just define matrix elements with the help of the following quantities
\begin{subequations}
\begin{align}
    &\mathcal{E}_\pm(T,\mu) = \frac{(d+a-2)B^{-\frac{d-2}{a}}}{2^{d-1}\pi^{\frac{d}{2}}d\Gamma(\frac{d}{2})}\left(k_BT\right)^{\frac{d-2}{a}+1} \nonumber\\
    & \qquad\times \Gamma\left(\frac{d-2}{a}+1\right)\Li_{\frac{d-2}{a}+1}(-z^{\pm1})~,\\
    &\mathcal{T}_\pm(T,\mu) = \pm\frac{(d+2a-2)B^{-\frac{d-2}{a}}}{2^{d-1}\pi^{\frac{d}{2}}d\Gamma(\frac{d}{2})}\left(k_BT\right)^{\frac{d-2}{a}+2}\nonumber\\
    &\qquad \times \Gamma\left(\frac{d-2}{a}+2\right)\Li_{\frac{d-2}{a}+2}(-z^{\pm1})- \mu\mathcal{E}_\pm(T,\mu)~,\\
    &\mathcal{S}_\pm(T,\mu) = \frac{(d+3a-2)B^{-\frac{d-2}{a}}}{2^{d-1}\pi^{\frac{d}{2}}d\Gamma(\frac{d}{2})}\left(k_BT\right)^{\frac{d-2}{a}+3}\nonumber\\
    &\qquad\times \Gamma\left(\frac{d-2}{a}+3\right)\Li_{\frac{d-2}{a}+3}(-z^{\pm1})\nonumber\\
    &\qquad\mp 2\mu\frac{(d+2a-2)B^{-\frac{d-2}{a}}}{2^{d-1}\pi^{\frac{d}{2}}d\Gamma(\frac{d}{2})}\left(k_BT\right)^{\frac{d-2}{a}+2} \nonumber\\
    &\qquad\times \Gamma\left(\frac{d-2}{a}+2\right)\Li_{\frac{d-2}{a}+2}(-z^{\pm1}) + \mu^2\mathcal{E}_\pm(T,\mu)~,
\end{align}
\end{subequations}
as
\begin{subequations}
\begin{align}
    \label{TransCoeffs}
    &\sigma = -e^2\frac{\mathcal{E}_+\left(\tau^{-1}_-+\tau^{-1}_\mathrm{dis} - \tau^{-1}_+\right) + \mathcal{E}_-\left(\tau^{-1}_+ +\tau^{-1}_\mathrm{dis} - \tau^{-1}_-\right)}{\tau^{-1}_\mathrm{dis}\left(\tau^{-1}_+ +\tau^{-1}_\mathrm{dis} + \tau^{-1}_-\right)}~, \\
    &\bar{\kappa} = -\frac{1}{T}\left[\frac{(\mathcal{S}_++\mathcal{S}_-)}{\tau^{-1}_\mathrm{dis}} + 2\frac{\mu\left(\tau^{-1}_-\mathcal{T}_- + \tau^{-1}_+\mathcal{T}_+\right)}{\tau^{-1}_\mathrm{dis}\left(\tau^{-1}_-+\tau^{-1}_\mathrm{dis} + \tau^{-1}_+\right)}\right]~,\\
    &\alpha = \frac{e}{T}\frac{\mathcal{T}_+\left(\tau^{-1}_-+\tau^{-1}_\mathrm{dis} - \tau^{-1}_+\right) + \mathcal{T}_-\left(\tau^{-1}_+ +\tau^{-1}_\mathrm{dis} - \tau^{-1}_-\right)}{\tau^{-1}_\mathrm{dis}\left(\tau^{-1}_+ + \tau^{-1}_\mathrm{dis} + \tau^{-1}_-\right)}~, \\
    &\bar{\alpha} = \frac{e}{T}\left[\frac{(\mathcal{T}_++    \mathcal{T}_-)}{\tau_\mathrm{dis}^{-1}} + 2\frac{\mu\left(\tau^{-1}_-\mathcal{E}_- + \tau^{-1}_+\mathcal{E}_+\right)}{\tau^{-1}_\mathrm{dis}\left(\tau^{-1}_-+\tau^{-1}_\mathrm{dis} + \tau^{-1}_+\right)}\right]~,
\end{align}
\end{subequations}
with the electric conductivity $\sigma$, the thermal one $\kappa = \bar\kappa - T\alpha^2/\sigma$ and the thermo-electric coefficients $\alpha$ and $\bar{\alpha}$, used to determine the Seebeck coefficient.
It should be noted that in this approach the relaxation times related to Coulomb scattering between electrons and holes and to effects of disorder enter all the observables and they ensure the well-definiteness of aforementioned coefficients.
In order to ensure the reciprocity of the system, Onsager's matrix should be symmetric and positive definite.
This condition $\alpha=\bar\alpha$ relates electron-electron relaxation times for different species $\tau_+$ and $\tau_-$ in a non-trivial way
\begin{equation}
\label{RelTimeRelation}
    \frac{1}{\tau_+}(\mathcal{T}_+ + \mu\mathcal{E}_+) = -\frac{1}{\tau_-}(\mathcal{T}_- + \mu\mathcal{E}_-)~.
\end{equation}
We then relate $\tau_+$ and $\tau_-$ to a characteristic relaxation time $\tau_0$, in the same fashion of~\cite{Pongsangangan2022HydrodynamicsTransport}.

\begin{figure}
  \centering
  \begin{subfigure}{0.96\columnwidth}
    \includegraphics[width=\linewidth]{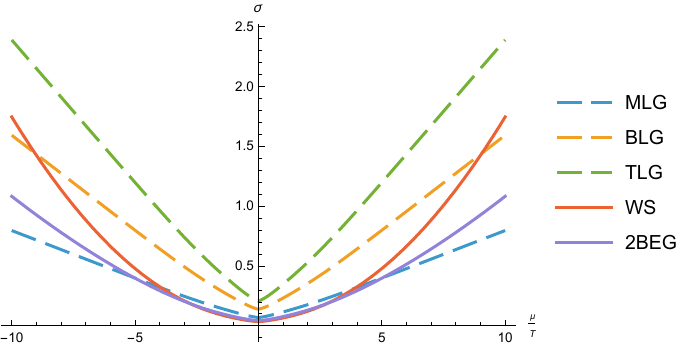}
    \caption{}
  \end{subfigure}
  \hfill
  \begin{subfigure}{0.96\columnwidth}
    \includegraphics[width=\linewidth]{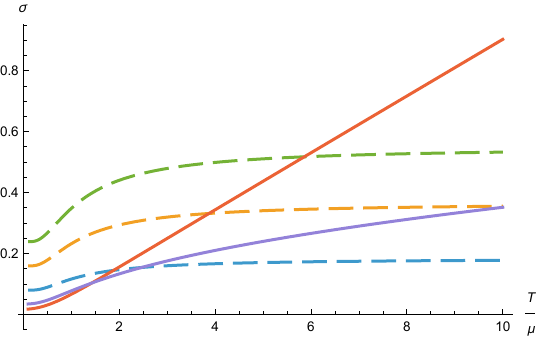}
    \caption{}
  \end{subfigure}
  \caption{
    We plotted here the value of electric conductivity $\sigma$ against the chemical potential in units of temperature $\mu/T$ (panel a) or temperature in units of chemical potential $T/\mu$ (panel b); we employ the same notation and color code as Fig.~(\ref{Fig:Eta}).  We notice that the thick lines grow parabolically against the linear growth of dashed ones in terms of chemical potential $\mu$; a similar conclusion can be obtained by studying the dependence on temperature $T$. We employed a constant $\tau_\mathrm{dis} = 1$ (in units of temperature or chemical potential, depending on the plot) and $\tau_0 = 0.6T\exp{(-|\mu|/T)}$.
  }
  \label{Fig:Sigma}
\end{figure}

\begin{figure}
  \centering
  \begin{subfigure}{0.96\columnwidth}
    \includegraphics[width=\linewidth]{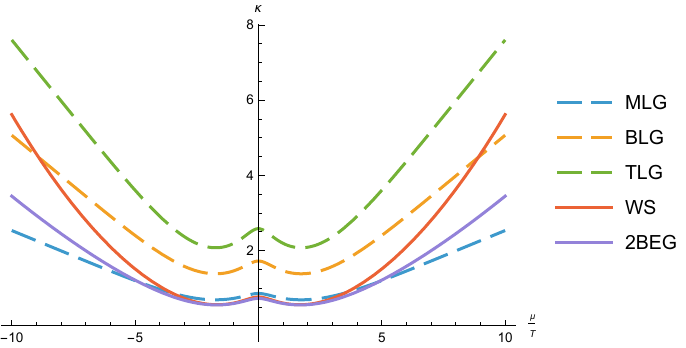}
    \caption{}
  \end{subfigure}
  \hfill
  \begin{subfigure}{0.96\columnwidth}
    \includegraphics[width=\linewidth]{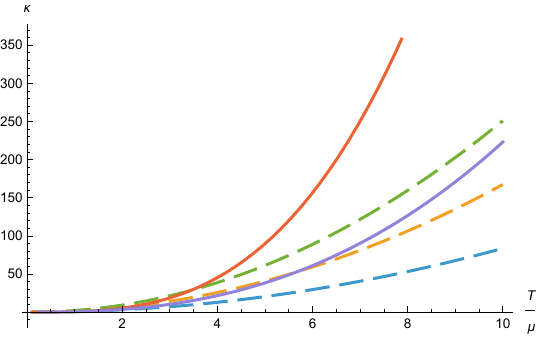}
    \caption{}
  \end{subfigure}
  \caption{
    We plotted here the value of thermal conductivity $\kappa$ against the chemical potential in units of temperature $\mu/T$ (panel a) or temperature in units of chemical potential $T/\mu$ (panel b); we employ the same notation and color code as Fig.~(\ref{Fig:Eta}). We notice again that the thick lines grow more rapidly in terms of chemical potential $\mu$ and temperature $T$. We employed a constant $\tau_\mathrm{dis} = 1$ (in units of temperature of chemical potential, depending on the plot) and $\tau_0 = 0.6T\exp{(-|\mu|/T)}$.
  }
  \label{Fig:Kappa}
\end{figure}

\begin{figure}
  \centering
  \begin{subfigure}{0.96\columnwidth}
    \includegraphics[width=\linewidth]{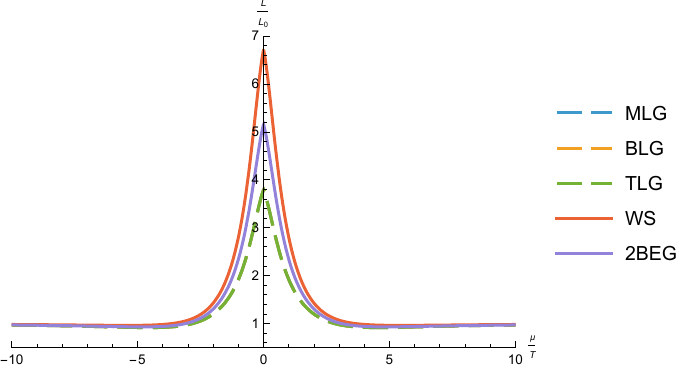}
    \caption{}
  \end{subfigure}
  \hfill
  \begin{subfigure}{0.96\columnwidth}
    \includegraphics[width=\linewidth]{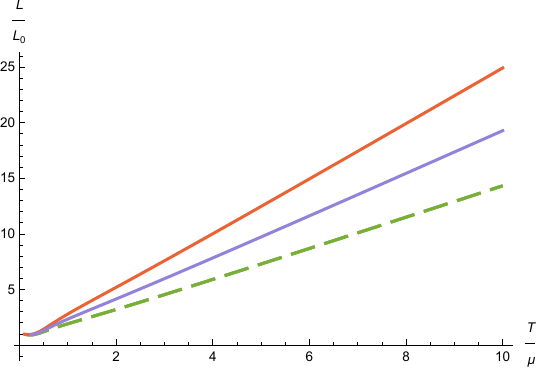}
    \caption{}
  \end{subfigure}
  \caption{
    We plotted here the value of Lorenz number $L/L_0$ against the chemical potential in units of temperature $\mu/T$ (panel a) or temperature in units of chemical potential $T/\mu$ (panel b); we employ the same notation and color code as Fig.~(\ref{Fig:Eta}). At the Dirac point the Lorenz numbers for different systems are maximal, while in the Fermi regime the Wiedemann-Franz law is restored for each of those. Lines for mono-, bi- and trilayer graphene are fully overlapping in these plots. We employed a constant $\tau_\mathrm{dis} = 1$ (in units of temperature of chemical potential, depending on the plot) and $\tau_0 = 0.6T\exp{(-|\mu|/T)}$.
  }
  \label{Fig:L}
\end{figure}

\begin{figure}
  \centering
  \begin{subfigure}{0.96\columnwidth}
    \includegraphics[width=\linewidth]{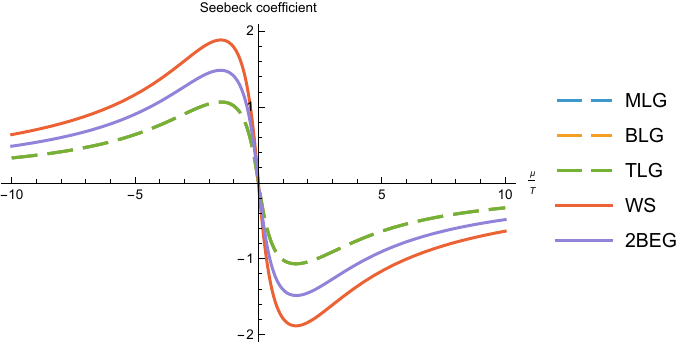}
    \caption{}
  \end{subfigure}
  \hfill
  \begin{subfigure}{0.96\columnwidth}
    \includegraphics[width=\linewidth]{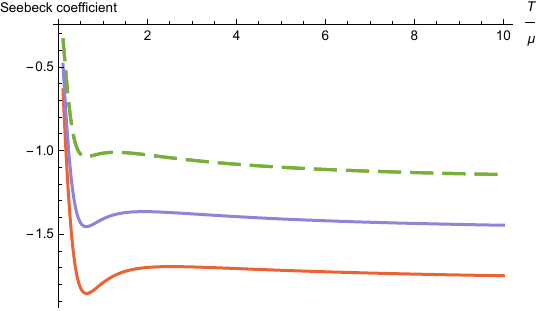}
    \caption{}
  \end{subfigure}
  \caption{We plotted here the value of Seebeck coefficient against the chemical potential in units of temperature $\mu/T$ (panel a) or temperature in units of chemical potential $T/\mu$ (panel b); we employ the same notation and color code as Fig.~(\ref{Fig:Eta}). The value of the coefficient is either positive or negative, being dependent on the nature of excitations transported; as a matter of facts, at charge neutrality it vanishes for any system considered. Lines for mono-, bi- and trilayer graphene are fully overlapping in these plots. We employed a constant $\tau_\mathrm{dis} = 1$ (in units of temperature of chemical potential, depending on the plot) and $\tau_0 = 0.6T\exp{(-|\mu|/T)}$.
  }
  \label{Fig:S}
\end{figure}

It has been known experimentally that metals follow the so-called Wiedemann-Franz law, which states that at low temperatures (compared to Fermi energy; i.e. in the Fermi limit) the ratio between the thermal conductivity and the electric one is a constant $L_0$, named Lorenz number, times the temperature; namely
\begin{equation}
    \label{LorenzNumb}
    L \equiv \frac{\kappa}{\sigma T} = L_0 = \frac{1}{3}\left(\frac{\pi k_B}{e}\right)^2.
\end{equation}
This fact can be easily explained in the picture of Drude model and equipartition of energy: those excitations which carry charge are the same carriers of energy and the latter scales linearly with temperature~\cite{Mahan-90}.
It is straightforward to understand why this does not hold for transport properties in the hydrodynamic regime: Hydrodynamic transport does not assume that the same modes contribute to charge transport and heat transport.
In fact, at the Dirac point of two-band systems, the Lorenz number reads
\begin{equation}
    \label{LorenzNumb2B}
    L = \frac{1}{e^2T^2}\frac{\mathcal{S}_++\mathcal{S}_-}{\mathcal{E}_++\mathcal{E}_-}\left(1 + \tau_\mathrm{dis}\frac{\tau_++\tau_-}{\tau_+\tau_-}\right)~\;.
\end{equation}
It is not a constant of temperature, instead it contains a explicit $T$ dependence, and it is not universal for any of the cases discussed here.
In this case, in the absence of disorder, $\tau_\mathrm{dis}\to\infty$, at charge neutrality we have $\tau_+=\tau_-$, and consequently the Lorenz number diverges. The intuition is this: The presence of an electric field moves electrons and holes inopposite directions allowing to establish a finite net electric current. A thermal gradient, moves electrons and holes in the same direction thereby exciting momentum. making the heat conductivity (\ref{LorenzNumb2B}) ill-defined.
Consequently, disorder has to be taken into account. 

On the other hand, if one computes the Lorenz number in the Fermi liquid limit ($\mathcal{S}_+ + \mathcal{S}_-\sim\mathcal{S}$, $\mathcal{E}_+ + \mathcal{E}_-\sim\mathcal{E}$ while $\tau_+\sim\tau_-\sim 0$), where effectively only one band contributes to transport and hydrodynamic properties of the electronic system, we recover the Wiedemann-Franz ratio, as shown in Fig.~\ref{Fig:L}.
This is possible since it is independent of the relaxation times as long as we consider the electronic system to be far from the Dirac point and disordered (diffusive).
In practice we have only one type of carrier which transports both charge and heat, while momentum is relaxed only via scattering with impurities, making the system transport properties in practice equivalent to those of regular Fermi liquid.

\begin{figure}
  \centering
  \begin{subfigure}{0.96\columnwidth}
    \includegraphics[width=\linewidth]{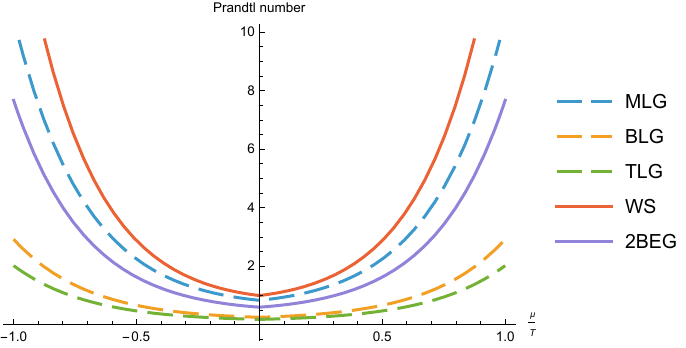}
    \caption{}
  \end{subfigure}
  \hfill
  \begin{subfigure}{0.96\columnwidth}
    \includegraphics[width=\linewidth]{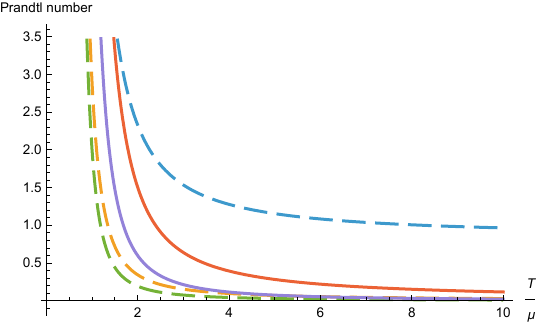}
    \caption{}
  \end{subfigure}
  \caption{
  We plotted here the value of Prandtl number against the chemical potential in units of temperature $\mu/T$ (panel a) or temperature in units of chemical potential $T/\mu$ (panel b); we employ the same notation and color code as Fig.~(\ref{Fig:Sigma}). The value of the coefficient is either positive or negative, being dependent on the nature of excitations transported; as a matter of facts, at charge neutrality it vanishes for any system considered. We employed a constant $\tau_\mathrm{dis} = 1$ (in units of temperature of chemical potential, depending on the plot) and $\tau_0 = 0.6T\exp{(-|\mu|/T)}$.
  }
  \label{Fig:PrandtlNumber}
\end{figure}

We can also compute the Prandtl number, a dimensionless quantity that considers the ratio between shear viscosity and heat conductivity.
Its importance lies in the fact that it fully describes the fluid state with no reference to any length scale, comparable to the Lorenz number for conducting materials.
At the Dirac point ($\mu=0$ and $\tau_+=\tau_-=\tau$) it reads
\begin{equation}
    \label{PrandtlNumberDirac}
    \mathrm{Pr}(T,\mu\ll k_BT) = \frac{c_p\eta}{\kappa} = \frac{2d}{a}\frac{\tau}{\tau_\mathrm{dis}}\frac{k_BP}{\mathcal{S}_+ + \mathcal{S}_-}~,
\end{equation}
and depends on the two time scales of the problem. Intuitively, the disorder dominated regime enhances momentum diffusion and showcases a Prandtl number that is significantly larger than $1$. The interaction dominated regime, on the other hand, shows large thermal diffusion and (\ref{PrandtlNumberDirac}) is way smaller than $1$.
If the interaction at the Dirac point is strong enough to lead the system to complete hydrodynamic behavior, i.e. $\tau\ll\tau_\mathrm{dis}$, then the clean limit is defined through a vanishing Prandtl number.

\section{Collective Modes}
\label{Collective Modes}
Strong electron-electron interactions, particularly the Coulomb interaction, play a crucial role in redistributing conserved quantities—particle number, momentum, and energy—across the system. While the short-range behavior of the Coulomb force governs local scattering processes, its long-range nature gives rise to collective screening effects. These are self-consistently mediated by the induced electric and magnetic fields that arise from the motion of the electron fluid~\cite{landau1981kinetictheory}.

In many materials, especially three-dimensional metals, these self-consistent fields can be treated within a quasistatic approximation due to the large energy gap associated with exciting longitudinal plasma oscillations, or plasmons~\cite{BohmPines2,BohmPines3,Pongsangangan2022HydrodynamicsModes}. Such excitations are inherently collective in character, involving coherent motion of many electrons, and are typically referred to as collective modes.

To understand the emergence of these collective modes within the hydrodynamic regime, particularly in single-band models interacting via the Coulomb potential, it is essential to analyze the underlying kinetic equation. This involves a detailed examination of the collision integral and the interaction kernel, which encodes the structure of electron-electron scattering and screening.

In the special case of linear energy bands where $a = 1$, it can be shown that the mass density behaves as $nm^\star = h / v_F^2 = (n\langle E\rangle + P)/v_F^2$. This provides a natural mechanism for closing the transport equations governing particle number, momentum, and energy, and thus leads to a consistent description of what are known as ``entropy waves''~\cite{landau1987fluid}, which are collective modes emerging from the energy transport equation. Similarly, for conventional parabolic bands, the presence of a constant effective mass enables the standard closure of these transport equations.

However, for a more general power-law dispersion where either $a \neq 2$ or $a \neq 1$, there is no straightforward way to express the mass density in terms of other macroscopic quantities across different carrier densities. Nonetheless, the thermodynamic relation $P = (a/d)\, n \langle E \rangle$ remains valid. To address this difficulty, we consider the Fermi liquid regime in the limit $\mu / T \to \infty$, allowing us to eliminate the chemical potential dependence of the hydrodynamic mass. Given that $\mu \propto n^{a/d}$, we can express the effective mass as $m^\star = m_0^\star n^{(2 - a)/d}$. Applying this, the linearization around a steady-state background with $n = n_0 + n_1$, $\vec{u} = \vec{u}_0 + \vec{u}_1$, and $n \langle E \rangle = n \langle E \rangle_0 + n \langle E \rangle_1$ yields
 
\begin{subequations}
\begin{align}
    \partial_t n_1+n_0\diver\vec{u}_1+ \vec{u}_0\cdot \grad n_1 = 0~, \\
    n_0m^\star_0\left(\ptderiv{\vec{u}_1}+(\vec{u}_0\cdot\grad)\vec{u}_1-\frac{2-a}{d} \vec{u}_0\diver\vec{u}_1\right)  \nonumber\\
    + \frac{a}{d}\grad n\langle E\rangle_1+ n_0 q \grad \phi_1 =0~,\\
    \partial_t n\langle E\rangle_1+\frac{d+a}{d}\vec{u}_0\cdot\grad n\langle E\rangle_1 \nonumber\\
    +\frac{d+a}{d}n\langle E\rangle_0\diver\vec{u_1}+n_0q \vec{u}_0  \cdot \grad \phi_1 = 0,
\end{align}
\end{subequations}
and the corresponding dispersion relation is given by 
\begin{equation}
    \begin{vmatrix}
    \omega -u_0k&-kn_0&0 \\
    - \frac{e^2}{\varepsilon m^\star_0}kK_d &\omega -\frac{d-2+a}{d}  u_0k &-\frac{a}{dn_0m^\star_0}k \\
    - u_0n_0\frac{e^2}{\varepsilon}kK_d &-\frac{d+a}{d}\epsilon_0k&\omega  -\frac{d+a}{d}u_0k 
    \end{vmatrix}=0~,
\end{equation}
where the factor $K_d$ is defined from the Coulomb kernel such that $\phi_1=\frac{q}{\varepsilon}n_1K_d(\vec{k})$. That is, in 3D $K_\text{3D}=1/k^2$ and in two-dimensional gated structures $K_\text{2D}=\tanh(d_0 k)/k\simeq 1/k$ as $d_0\to\infty$. 

So, within this framework, one can retrieve a low-frequency mode with $\omega=u_0k$ that entails no perturbation in the velocity field, i.e. $u_1=0$, and 
\begin{equation}
  \frac{a}{d} n\langle E\rangle_1=-n_0  \frac{e^2}{\varepsilon}K_d(k)n_1~,
\end{equation}
so that the energy, or equivalently pressure or enthalpy, oscillates in contrast to the number density while the perturbation is carried along the flow. Such a mode is, in all respects, analogous to the entropy-vortex waves of classical fluids~\cite{landau1987fluid}. 

Additionally, the modes where $n\langle E\rangle_1=0$ lead to plasmons with dispersion
\begin{equation}
    \omega=\left(1+\frac{a-2}{2d}\right)u_0k\pm  \sqrt{k^2K_d(k)\frac{e^2n_0}{m_0^\star\varepsilon}+u_0^2k^2\left(\frac{a-2}{2d}\right)^2}~,
\end{equation}
and it agrees with the predictions that plasmonic modes are gapped in 3D systems, while they are not for 2D ones and their dispersion scales as $k^{1/2}$, without depending on the details of (\ref{eq:powerlaw}). Additionally, once again the breaking of Galilean boosts for $a\neq 2$ is made clear by the prefactors of $u_0$. 

\section{Conclusions}
\label{Conclusions}

In this work, we studied the hydrodynamic equations of two-band electronic systems with generic dispersion relation (\ref{eq:powerlaw}) and conserving electron-electron interaction, both at Euler and Navier-Stokes regimes.
We consider generic dimensionality of the system $d$ to include bi- and tri-dimensional materials in the framework.
Our approach is based on the kinetic theory and Boltzmann equation formalism, allowing to derive Euler and Navier-Stokes equations.
These system of equations can be closed in two specific configurations of the two-band system, i.e. in the vicinity of Dirac point $\mu\ll k_BT$ and at the Fermi limit $\mu\gg k_BT$; such points are also relevant under the physical perspective for the reason that conserving interaction is expected to dominate over disorder and other dissipative effects across one of them.
Moreover, systems with power-law exponent $a\neq 2$ break Galilean invariance of the macroscopic system and we show the consequences on conservation of momentum and transport of conserved quantities.
Specifically, systems with linear dispersion as graphene or Weyl semimetals do not recover Lorentz invariance, but the connection between momentum and charge density is translated into the one between momentum and energy, allowing exact closure of hydrodynamic equations also far from Dirac or Fermi regimes.
In presence of dissipation but in the clean limit, we are able to compute the viscosity of these systems, retrieving some exact results already contained in the literature for linear and quadratic dispersions.
When disorder is included, we can also define transport coefficients at the Dirac point since they are particularly sensitive to momentum relaxation.
We quantify the deviation from the Wiedemann-Franz law and the ratio between momentum versus heat diffusivity of the quantum liquid: we observe strong violations of the assumptions underlying Drude-like models, such as the equivalence between charge and heat carriers, in particular when the system's configuration is in the vicinity of the Dirac point.
Finally, we show the emergence of collective plasmonic modes in the hydrodynamic regime.
Their impact on transport properties of graphene and Weyl semimetals has been studied in~\cite{Pongsangangan2023TheGraphene,Bernabeu25}, but a complete picture will be subject of a forthcoming publication.

\section{Acknowledgments}
\label{Acknowledgments}

We acknowledge Kitinan Pongsangangan and Hugo Terças for useful discussion.
E.D. and P.C. contributed equally to the present work.
P.C. acknowledges support from the ERC grant no.~2023-StG-101117025, FluMAB.
This work is part of the D-ITP consortium, a program of
the Dutch Research Council (NWO) that is funded by the Dutch Ministry of Education,
Culture and Science (OCW).

\bibliography{BIG-Bibliography}

\appendix

\section{Integration over isotropic multi-dimensional momentum space}
\label{Appendix: Integration over isotropic multi-dimensional momentum space}
    We derive here some useful properties of the following tensors in the $d$-dimensional space
    \begin{equation}
        \label{tensor}
        T^{a_1\dots a_n} = \int_{\mathbb{R}^d}\frac{\dd^d \kk}{(2\pi)^d} (\kk)_{a_1}\cdots(\kk)_{a_n} g(k)~,
    \end{equation}
    where $g$ is just a function of the modulus of the vector $\kk$.
    It is straightforward to derive the invariance of the integral measure $\int_{\mathbb{R}^d}\frac{\dd^d \kk}{(2\pi)^d}$ under rotation of $\pi$ degrees of one axis $(\kk)_{a}\to - (\kk)_{a}$ and of rotation between different axis $(\kk)_{a}\leftrightarrow (\kk)_{b}$; also the function $g$ is not changed by those transformations.
    Hence, the tensor (\ref{tensor}) enjoys the following properties
    \begin{enumerate}
        \item It is fully symmetric: $T^{a_1\dots,a_{i+1},a_{i},\dots a_n} = T^{a_1\dots a_n}$ ,
        \item It has vanishing elements for odd $n$ ,
        \item For even $n$, it has vanishing off-diagonal elements: if $a\neq b$, $T^{aba_1a_1\dots a_na_n} = 0$ ,
        \item All diagonal elements are equal.
    \end{enumerate}
    We can specify them for the cases $n=2$ and $n=4$, which are of interest in our coming derivations.
    The tensor $T^{ab}$ is then proportional to Kronecker delta as long as all its off-diagonal elements are vanishing and the diagonal elements are all equal
    \begin{equation}
        T^{ab} = T\delta^{ab}~.
    \end{equation}
    Now the trace of $T$ reads
    \begin{equation}
        \mathrm{Tr}(T) = \int_{\mathbb{R}^d}\frac{\dd^d \kk}{(2\pi)^d}k^2 g(k) = Td~,
    \end{equation}
    and it allows to derive the exact value of $T$.

    The rank-$4$ tensor has indeed more structure in it: it is forced by the aforementioned constraints to be vanishing when two indices are not equal to any other and we can reduce to problem to studying
    \begin{equation}
        \bar{T}^{ab} = T^{aabb} = \int_{\mathbb{R}^d}\frac{\dd^d \kk}{(2\pi)^d} (\kk)^2_{a}(\kk)^2_{b} g(k) = \bar{T}^{ba}~,
    \end{equation}
    that is a symmetric tensor with off-diagonal terms equal to $\bar{T}$ and diagonal ones to be $\bar{T} + \Delta$; then
    \begin{eqnarray}
        &\sum_{a,b}^n \bar{T}^{ab} = \int_{\mathbb{R}^d}\frac{\dd^d \kk}{(2\pi)^d} k^4 g(k) = d^2\bar{T} + d\Delta~, \\
        &\mathrm{Tr}(\bar{T}) = \int_{\mathbb{R}^d}\frac{\dd^d \kk}{(2\pi)^d} \sum_{a=1}^n(\kk)^4_{a} g(k) = d \Delta~.
    \end{eqnarray}
    Then, $\Delta = \mathrm{Tr}(\bar{T})/d$ and $\bar{T} = \left[\sum_{a,b}^n \bar{T}^{ab} - \mathrm{Tr}(\bar{T})\right]/d^2$.

\section{Derivation of Euler equations}
\label{Appendix: Derivation of Euler equations}

    We are now in the position of deriving Euler equations for a model that conserves particle number, momentum and energy.
	We always start from the collisionless Boltzmann equation, multiply for the conserved microscopic quantity, which is a function of momentum/wave vector $\kk$, and integrate over $d$-dimensional space of momenta.
	in this way we retrieve the usual continuity equations
	\begin{equation}
		\partial_t q - \vec{\nabla}\cdot\vec{j} = 0~,
	\end{equation}
	which is a tensor equation in case $q$ is a vector, or an higher-rank tensor.
	
	\subsection{Charge equation}
	It this case the conserved charge current, we can derive it by computing the contributions by different species, given by the following integrals
	\begin{align}
		&\vec{j}_\pm = \pm\int_{\mathbb{R}^d}\frac{\dd \kk^d}{(2\pi)^d}\vec{\nabla}\omega f^{(\pm)}(\omega)  \\
        &=-\int_{\mathbb{R}^d}\frac{\dd \kk^d}{(2\pi)^d} (\uu\cdot\kk)\vec{\nabla}\omega\frac{\partial f^{(\pm)}_\mathrm{eq}}{\partial\omega}(\omega) + O(|\uu|^2)~. \nonumber
	\end{align}
	This integral can either be twisted into the tensor form $I_{ij} = I\delta_{ij}$ by isotropy considerations or by direct integration by parts
	\begin{eqnarray}
		& \vec{j}_\pm = n_\pm\uu~,
	\end{eqnarray}
	where we retrieve the usual relation between velocity field $\uu$ and charge current.
    The charge current is defined as
    \begin{equation}
        \vec{j}_c = \vec{j}_+ + \vec{j}_-~.
    \end{equation}
	This allows us to write down the first Euler equation for charge density
	\begin{equation}
		\partial_t n_{c} - \vec{\nabla}\cdot(n_{c}\uu) = 0 ~.
	\end{equation}
	This relation is valid beyond the hydrodynamic regime and just states the conservation of charge for electronic systems.
	
	\subsection{Momentum equation}
	Being momentum $\pp$ a vector, the related conserved current is the tensor $\vec{\vec{\Pi}}$, which we are going to express in terms of its matrix elements $\Pi_{ij,\pm}$
	\begin{equation}
		\Pi_{ij,\pm} = \pm\int_{\mathbb{R}^d}\frac{\dd^d \kk}{(2\pi)^d}(\kk)_i\left(\vec{\nabla}\omega\right)_j f^{(\pm)}(\omega)~.
	\end{equation}
	Now, in order to get the correct convective term for the final equation, we need to expand the perturbed distribution function up to the second order.
	Although, let's start from the zero order
	\begin{equation}
		\pm  aB\int_{\mathbb{R}^d}\frac{\dd^d \kk}{(2\pi)^d} k^{a-2}(\kk)_i(\kk)_j f^{(\pm)}_\mathrm{eq}(\omega)~,
	\end{equation}
	which is again proportional to Kronecker delta and the scalar prefactor is as a matter of facts the pressure experienced by the quantum fluid itself
	\begin{equation}
		P = \frac{aB}{d}\int_{\mathbb{R}^d}\frac{\dd^d \kk}{(2\pi)^d} k^{a} \left(f^{(+)}_\mathrm{eq}(\omega) - f^{(-)}_\mathrm{eq}(\omega)\right) = \frac{a}{d}\varepsilon_c~,
	\end{equation}
	and it is indeed proportional to the internal imbalance of energy of the system itself.
	
	Second order needs a bit more of effort
	\begin{equation}
    \label{ViscTens0}
		 \pm\frac{aB}{2} (\uu)_n(\uu)_m\int_{\mathbb{R}^d}\frac{\dd^d \kk}{(2\pi)^d} k^{a-2}(\kk)_i(\kk)_j (\kk)_n (\kk)_m \frac{\partial^2f^{(\pm)}_\mathrm{eq}}{\partial\omega^2}(\omega)~,
	\end{equation}
	since we have to deal with integrating a fourth order rank tensor.
    Here however isotropy and rotational invariance of the system come at rescue and allow us to write it as a product of Kronecker deltas and a scalar in front $\Lambda(\delta_{ij}\delta_{nm} + \delta_{in}\delta_{jm} + \delta_{im}\delta_{jn}) + \sigma\delta_{ijmn}$.
    $\Lambda$ can be connected with the mass density as
	\begin{align}
		\Lambda = c_2(d,\alpha) \gamma(d,\alpha)\rho_\pm~, \nonumber\\
		c_2(d,\alpha) =  \frac{1}{d(d+2)}~,
	\end{align}
	and introducing the new dressing coefficient. Write down the derivation of tensor structures by isotropy and rotational invariance here.
	In fact, if we write down the second Euler equation, which is the only one which trivially relates the state functions to their conservation laws, we have
	\begin{equation}
		\partial_t(\rho_c\uu) - c_2(d,\alpha,\beta)\rho_c(\uu\cdot\vec{\nabla})\uu = -\vec{\nabla}P~,
	\end{equation}
	where we have neglected terms of order $|\uu|^2$.
	
	\subsection{Energy equation}
	At last, we go back to energy conservation and the energy current for the species is just:
	\begin{multline}
		\vec{j}^\varepsilon_\pm = \int_{\mathbb{R}^d}\frac{\dd^d \kk}{(2\pi)^d}\omega\vec{\nabla}\omega f^{(\pm)}(\omega)  \\
        = - aB^2\int_{\mathbb{R}^d}\frac{\dd^d \kk}{(2\pi)^d} k^{2a-2}\kk(\uu\cdot\kk)\frac{\partial f^{(\pm)}_\mathrm{eq}}{\partial\omega}(\omega) + O(|\uu|^2)~.
	\end{multline}
	We can prove the charge current can be written as $W\uu$ using the usual isotropy argument and the related tensor representation, where $W$ is the enthalpy of the system:
	\begin{align}
		& W = -\frac{aB^2}{d}\int_{\mathbb{R}^d}\frac{\dd^d \kk}{(2\pi)^d} k^{2a}\frac{\partial}{\partial\omega}\left(f^{(+)}_\mathrm{eq}(\omega)-f^{(-)}_\mathrm{eq}(\omega)\right)\nonumber\\
        &\quad = \left(1 + \frac{a}{d}\right)\varepsilon_c ~,
	\end{align}
	and the final dressing coefficient shows up; we retrieve the thermodynamic relation between enthalpy, energy and pressure $W = \varepsilon_\mathrm{imb} + P$.
	Then the third Euler equation, involving conservation of energy, is written as
	\begin{equation}
		\partial_t\varepsilon_c + \left(1 + \frac{a}{d}\right)\vec{\nabla}\cdot(\varepsilon_c\uu) = 0~.
	\end{equation}

\section{Breaking of Galilean invariance}
\label{Appendix: Breaking of Galilean invariance}

One consequence of the variable mass element is the loss of boost invariance, i.e. the system is neither Galilean nor Lorentz invariant. Let us see why this is the case. We established in Eq.\eqref{eq:generalmomentum} that, for a variable hydrodynamic mass element (following a power law), the convective part of the momentum equation acquires an additional term $\sim \vec{u}\diver\vec{u}$. To understand its effect on the symmetry properties of the fluid equations we consider the system, comprised of equations $\Delta_1$ and $\Delta_2$, to be uniform along the $y$ direction, arriving at
\begin{subequations}
\begin{gather}
    \Delta_1\triangleq\partial_tu+\alpha u\partial_x u+\partial_x n=0~,\\
    \Delta_2\triangleq\partial_tn+u\partial_x n+n\partial_x u=0~,
\end{gather}\label{eq:shallowSymm}%
\end{subequations}%
which we can categorize as generalized shallow water equations. Such a system captures the behaviour and symmetries of our hydrodynamic models up to rotations. 

Determining the allowed symmetries of a set of PDEs might not be self-evident. Here we give an abridged account of the general procedure applied to Eqs.\eqref{eq:shallowSymm}; for further details refer to  \cite{Hydon2000SymmetryEquations}.
On a system governed by some set of partial differential equations, $\Sigma=\{\Delta_1,\Delta_2,\ldots,\Delta_\ell\}$, let the independent variables $\Vec{x}\in\Omega\subset\mathbb{R}^p$ and the dependent variables $\Vec{u}\in U \subset\mathbb{R}^q$. Moreover, let $U_k$ be the space of all $k$-th order derivatives of $\Vec{u}$, with respect to $\Vec{x}$. With this setting, we define the jet space~\cite{Olver1986ApplicationsEquations} of order $n$ as
\begin{equation}
    M^{(n)}=\Omega\times U \times U_1\times \cdots\times U_n~,
\end{equation}%
that is also known as the $n$-th prolongation space of the space of variables $M =\Omega\times U $ where the solutions of $\Delta=0$ live. 
A continuous group $(G,\cdot)$ acting on the space of variables $M$ is a symmetry of the differential equation $\Delta=0$ if the solution submanifold 
\begin{equation}
    \mathcal{S}_\Delta=\{(\Vec{x},\Vec{u},\partial_\Vec{x}{\Vec{u}},\partial_\Vec{x}^2\Vec{u}, \dots,\partial_\Vec{x}^n\Vec{u}):\Delta=0\}\subset M^{(n)}
\end{equation}
remains invariant under its action, i.e. if $G\cdot\mathcal{S}_\Delta\subseteq \mathcal{S}_\Delta$. 

To find the group from the PDE we start by looking for basis vectors of the associated Lie algebra $\mathfrak g :G=\exp(\mathfrak{g})$ resorting to the following theorem~\cite{Olver1986ApplicationsEquations,HYDON2000HowEquations,Hydon2000SymmetryEquations}:

\bigskip
\emph{
A Lie group $G$, acting on $M \cong\mathbb{R}^p\times\mathbb{R}^q $, is a symmetry group of the system, $\Sigma$, of $\ell$ $n$-th order differential equations $\Delta_i(\Vec{x},\Vec{u}^{(n)})=0$ if and only if:
\begin{equation}
    \pro{n}\Vec{v}(\Delta_i)\Big|_\Sigma=0\quad \text{for } i =1,\ldots,\ell\label{eq:theoremLieProlongationsMain}
\end{equation}%
for every generator $\Vec{v}\in \mathfrak{g}$ of $G$.%
}\bigskip

\noindent There $\pro{n}\Vec{v}$ stands for the prolongation of the vector $\Vec{v}$, defined as 
$   \pro{n}\Vec{v}=\Vec{v} +\sum_{k=1}^q\sum_{J} \widetilde\eta^{J}_k(\Vec{x},\Vec{u}^{(n)})\pderiv{}{u_{J}^k}$
where the coefficients $\widetilde\eta^J$ are computed iteratively by the total derivatives 
$\widetilde\eta^{i}_k = D_i(\eta^k)- \sum_{l=1}^p u^k_l D_i(\xi^l)$.
 
Applying this theorem to Eqs.\eqref{eq:shallowSymm} for an infinitesimal generator of the form
$\Vec{v}=\tau(t,x,u,n)\partial_t+\xi(t,x,u,n)\partial_x+\eta(t,x,u,n)\partial_u+\phi(t,x,u,n)\partial_n$ which completely defines a one-parameter group on our variables, and equating the monomials of $M^{(n)}$, yields the overdetermined  set of equations 
\begin{subequations}
\begin{gather}
-(1-\alpha)u \eta_n+\phi_n-\eta_u+\tau_t\nonumber \\\qquad\qquad -\xi_x+(1+\alpha)u \tau_x=0~,\\ 
\phi+(1-\alpha)u\phi_u+\nonumber\\n\left( -\phi_n+ \eta_u+ \tau_t-\xi_x+ (1+\alpha)u \tau_x \right)=0~,\\ 
\eta+n \eta_n+(n+u^2)\tau_x+u\tau_t\nonumber\\\qquad -u\xi_x -\phi_u-\xi_t=0~,    \\
a\eta-n \eta_n+\left(n+\alpha^2u^2\right)\tau_x+\nonumber\\\qquad \alpha u\tau_t-\alpha u\xi_x+\phi_u-\xi_t=0~, \\ 
\xi_n-\alpha u\tau_n+\tau_u=0~,\\ 
\xi_u+n \tau_n-u \tau_u=0~,\\ 
\eta_t+\alpha u \eta_x+\phi_x=0~,\\ 
\phi_t+n \eta_x+u \phi_x=0~,
\end{gather}\label{eq:determining_equations}%
\end{subequations}%
where we have used the shorthand $\partial_x\tau \triangleq\tau_x$ and so forth for all the variables. 
It is already blatant that the particular case of $\alpha=1$ renders this set less convoluted.
Solving Eqs.\eqref{eq:determining_equations} for an arbitrary value of $\alpha\neq1$ shows that the Lie algebra, $\mathfrak{g}=\mathfrak{g}_2\oplus\mathfrak{g}_\infty$, is spanned by the vectors 
\begin{equation}
    \begin{gathered}
    \Vec{v}_1=t\partial_t +x\partial_x ~,\quad \Vec{v}_2 = 2n\partial_n+u\partial_u-t\partial_t~,\\
    \Vec{v}_{\infty} = F^1(u,h)\partial_t+F^2(u,h)\partial_x 
    \end{gathered}
\end{equation}
of a two-dimensional Abelian subalgebra $\mathfrak{g}_{2}=\langle\Vec{v}_1,\Vec{v}_2\rangle$ of dilations; and the infinite-dimensional subalgebra $\mathfrak{g}_{\infty}\ni \Vec{v}_{\infty}$ where $F^1$ and $F^2$ obey $F^2_n-u F^1_n+F^1_u=0,\, F^2_u+n F^1_n-u F^1_u=0$, note that the temporal or spatial translations  (correspondingly $\Vec{v}=\partial_t$ and $\Vec{v}=\partial_x$) can be seen as a trivial case of the latter when $F^1=1\wedge F^2=0$ and $F^1=0\wedge F^2=1$, respectively. Moreover, it is $\mathfrak{g}_\infty$, or rather the structure of its vectors, that guarantees the applicability of a hodograph transformation to the PDE system~\cite{Dorodnitsyn2020ShallowModels}. But from a physics perspective, only dilations and translations are present.
However, in the particular case of $\alpha=1$ and only for that, corresponding to the case of parabolic dispersion relations and thus to regular hydrodynamics, the Lie algebra is extended by the vectors
\begin{equation}
  \begin{gathered} 
    \Vec{v}_3= t\partial_x+\partial_u~,  \\
    \Vec{v}_4 = 4nu\partial_n+(4n+u^2)\partial_u+(2x-6ut)\partial_t+3t(2n-u^2)\partial_x
    \end{gathered}
\end{equation}
so that now the subalgebra is $\mathfrak{g}_{4}=\langle\Vec{v}_1,\Vec{v}_2,\Vec{v}_3,\Vec{v}_4\rangle$, which now includes the generator for Galilean boosts. Yet, let us stress that these boosts do not originate from a group contraction of a Poincaré-like group and that the algebra $\mathfrak{g}_{2}\subset\mathfrak{g}_{4}$ contains all the possible polynomial infinitesimal generators. Furthermore, it is now evident that even in the case of relativistic particles, i.e. with a linear dispersion as graphene electrons, the mean-field hydrodynamic description does not inherit the Lorentz group structure. In this way, this discussion, and the conclusion that the invariant action of boosts is not possible for $\alpha\neq1$ because $m^\star\neq \rm{const.}$

\section{Derivation of shear viscosity}
\label{Appendix: Derivation of shear viscosity}

    In this section we derive explicitly the shear viscosity for $d=2$, $d=3$ and $d>3$ systems.
    An assumption we employ to make connection with our result with total thermodynamics quantities is that $\tau_+ = \tau_- = \tau$; the general result for $\tau_+\neq\tau_-$ can be trivially obtained with the same method.
    We start from the following expressions for the stress-momentum tensor and its contributions $\vec{\widetilde{\Pi}}_\pm = \vec{\widetilde{\Pi}}^{(1)}_\pm + \vec{\widetilde{\Pi}}^{(2)}_\pm,$ from the scalar product $\grad\cdot\uu$ and the external one $\grad_i\uu_j$
    \begin{subequations}
        \begin{align}
        &\vec{\widetilde{\Pi}}^{(1)}_{\pm,ij} = \\ &\mp\tau\left[\int\frac{\dd^d\mathbf{k}}{(2\pi)^d}v_ik_jk_lv_n f^{(\pm)}_\mathrm{eq}(k)\left(1-f^{(\pm)}_\mathrm{eq}(k)\right)\right]\grad_n\uu_l~,\nonumber\\
        &\vec{\widetilde{\Pi}}^{(2)}_{\pm,ij} = \\
        &\pm\frac{\tau}{d}\left[\int\frac{\dd^d\mathbf{k}}{(2\pi)^d}v_ik_jk_lv_l f^{(\pm)}_\mathrm{eq}(k)\left(1-f^{(\pm)}_\mathrm{eq}(k)\right)\right]\grad\cdot\uu~.\nonumber
    \end{align}
    \end{subequations}
    By taking the trace one can directly see that the bulk viscosity vanishes.

    In order to retrieve the shear viscosity coefficient, we just compute the element $\vec{\widetilde{\Pi}}^{(1)}_{\pm,12}$.
    Isotropy ensures that it is the same computing any of them, as long as they are not diagonal, which are vanishing.
    To such element, only the first part of the stress-momentum tensor contributes, since the second cannot allow for off-diagonal pieces, and it reads
    \begin{multline}
        \vec{\widetilde{\Pi}}^{(1)}_{\pm,12} =\mp\tau\grad_n\uu_l\times \\
        \left[\int\frac{\dd^d\mathbf{k}}{(2\pi)^d}k^{2a-4}k_1k_2k_lk_n f^{(\pm)}_\mathrm{eq}(k)\left(1-f^{(\pm)}_\mathrm{eq}(k)\right)\right]~,
    \end{multline}
    which we now rewrite as a product with a tensor $\vec{\widetilde{\Pi}}^{(1)}_{+,12} + \vec{\widetilde{\Pi}}^{(1)}_{-,12} = -2\tau T_{12ln}\grad_n\uu_l\equiv\eta(\grad_2\uu_1+\grad_1\uu_2)$, in the same spirit of Appendix \ref{Appendix: Integration over isotropic multi-dimensional momentum space}.
    We can now evaluate $\eta$ using the same techniques developed above, but we need to specify the dimension for $d=2$ and $d=3$ in order to express the integration measure.
    For $d>3$ a generic result can be found.
    For each of these cases, we can identify a geometric factor that multiplies the integral over the modulus of $\kk$ and the prefactor $\Omega_d$
    \begin{eqnarray}
        &\mathcal{G}_2 = 1/8 ~,\nonumber\\
        &\mathcal{G}_3 = 1/15 ~,\nonumber\\
        &\mathcal{G}_d = \Gamma(d)\Gamma(d-3)/\left[2^{d-2}d(d-3)\Gamma^2\left(d/2\right)\Gamma^2\left((d-3)/2\right)\right].\nonumber
    \end{eqnarray}
    The contribution from radial integration does not depend on these details and it gives $d(d+a)P$, leading to the general result
    \begin{equation}
        \eta = -2\tau \mathcal{G}_d d(d+a)P~,
    \end{equation}
    which agrees with the expressions on the main text.
\end{document}